\documentclass[a4paper]{article}

\usepackage{INTERSPEECH2020}
\usepackage{subcaption}
\usepackage{graphicx}
\usepackage{lipsum}
\usepackage{xcolor}
\title{The NTU-AISG Text-to-speech System for Blizzard Challenge 2020}
\name{
    Haobo Zhang$^1$,
    Tingzhi Mao$^1$,
    Haihua Xu$^2$,
    Hao Huang$^1$
    }
\address{
  $^1$School of Information Science and Engineering, Xinjiang University, Urumqi, China\\
  $^2$School of Computer Science and Engineering, Nanyang Technological University, Singapore
  }
\email{\{hallboo147,tingzhimao,hhx502\}@gmail.com}
\begin{document}

\maketitle
\begin{abstract}
We report our NTU-AISG Text-to-speech (TTS) entry systems for the Blizzard Challenge 2020 in this paper. There are two TTS tasks in this year challenge, one is a Mandarin TTS task, the other is a Shanghai dialect TTS task.  We have participated both. One of the main challenges is to build TTS systems with low-resource constraints, particularly for the case of Shanghai dialect, of which about \texttt{three} hours data are available to participants. To overcome the constraint, we  adopt 
an average-speaker modeling method. That is, we first employ external Mandarin data to train both End-to-end acoustic model and  WaveNet vocoder, then we use Shanghai dialect to tune the acoustic model and WaveNet vocoder respectively.    Apart from this, we have no Shanghai dialect lexicon despite syllable transcripts are provided for the training data. Since we 
are not sure if similar syllable transcripts are provided for the evaluation data during training stage,
we use Mandarin lexicon for Shanghai dialect instead. With the letter, as decomposed from the corresponding Mandarin syllable, as input, though the naturalness and original speaker similarity of the synthesized speech is good, subjective evaluation results indicate
the intelligibility of the synthesized speech is deeply undermined for the Shanghai dialect TTS system.

% it is found a sensible Shanghai dialect TTS system can be yielded with the help of transfer learning method as mentioned. 

% This paper introduces a Text-to-speech (TTS) system developed by NTU-AISG for the Blizzard Challenge 2020.
% There are two tasks in this year. We build a Mandarin and a Shanghainese system in the same framework.
% In our system, The Front-end converts the text to the Mandarin syllable sequence as the linguistic feature. Then the Back-end based on Tacotron2 encode the syllable sequence and generate the Fbank acoustic feature from the linguistic feature encoded.
% Finally, a WaveNet-based neural vocoder was used to generate speech waveforms from the Fbank acoustic features.
% In the task2 (SS1), Shanghainese voice is only three hours.
% The small size of the dataset meant that it was not possible to develop a Shanghainese system to build a high-quality voice.
% So investigated a method based on transfer learning to build Shanghainese voice from limited data.


% The first task (MH1) is build mandarin voice from about 9.5 hours of highly expressive Mandarin speech data. The second task (SS1) is build Shanghainese voice just from about three hours speech.
\end{abstract}

\section{Introduction}
As Artificial Intelligence (AI) technology is widely applied in human daily life nowadays, human-machine interaction, such as human-machine speech-based dialogue interaction, has become increasingly important. Text-to-speech (TTS) is one of many key components realizing 
a better human-machine interaction experience.
Over decades, TTS technology has been developed significantly, from the earliest hand-crafted wave unit concatenation method~\cite{hunt1996unit,campbell1997prosody} to the present machine-learning-based End-to-end method~\cite{wang2017tacotron,shen2018natural,donahue2020end,ren2019fastspeech}, thanks to the advent of the deep neural network technologies~\cite{chorowski2015attention,oord2016wavenet,paine2016fast}. Previously, it is not a trivial work to build a TTS system, since one needs expertise for different components.

Though we have achieved remarkable progress on TTS technology development, we are still far from perfect in many scenarios. For large high quality speech training data that are recorded by professional speaker from sound recording studio, we are almost done with perfect text-to-speech quality~\cite{shen2018natural}. For limited speech data that are recorded with inconsistent acoustic environment, the quality of the synthesised speech are yet to be improved. Besides, for highly expressive speech, such as speech enriched with emotion, present methods still have long way to go for final mature application. Additionally,
majority of present TTS systems only perform monolingual speech synthesis. As globalization trends are intensified, code-switching~\cite{E2E2020CSTTS,cao2019end}  based multilingual speech synthesis is also worth our efforts.
Last but not least, state-of-the-art TTS systems are often built with deep neural networks, which can have up to several million parameters that requires big device storage, as well as extensive computational resource. Consequently, how to develop small foot print  device-based TTS system is also drawing our attention~\cite{kalchbrenner2018efficient,tobing2020efficient,valin2019lpcnet}.

To address the above-mentioned challenges and the beyond, as well as to explore the new frontier of the novel Text-to-speech method, Blizzard Challenge workshop has been initiated since 2005~\cite{black2005blizzard}. It is drawing word-wide attention and participation from both research and industrial communities, expediting the process of research and development for TTS system.  

There are two sub-tasks in this year challenge. The first one is a Mandarin TTS task, and the second one is a Shanghai dialect TTS task. Training data are \texttt{9.5} and \texttt{2.9} hours respectively. We participated both tasks. The main challenge comes from that the second task belongs to a low-resource problem.
Our two TTS systems are built with Tacotron2~\cite{shen2018natural} plus WaveNet~\cite{oord2016wavenet} . Specifically, we use the End-to-end approach to front-end acoustic modeling. It realizes text to acoustic feature estimation. After that, a WaveNet conditioned on the learned acoustic features is employed as vocoder to synthesize the waveform.

The main efforts lie in how to build a TTS system  under such a low-resource condition, particularly for the Shanghai dialect case. Since we have no knowledge and resource about Shanghai dialect, except for that it shares the same character sets with the Mandarin language, we leverage both Mandarin speech data and lexicon to solve the problem.
In practice, we use Mandarin speech data to train a TTS system, then we tune the TTS system using Shanghai dialect. Since we realized it is obvious wrong to let the two languages share the same lexicon, we use letter instead, which is decomposed from Mandarin syllable, as input to the Tacotron models~\footnote{After system submission, we realized our method has severe defect. Not only can the syllables of the two languages not be shared, but also the letter of the syllable cannot be shared. Heuristically, it should be better to add language-dependent identifiers to each letter of the syllable, differentiating the two languages.}. After evaluation results come out, we found out the intelligibility of our synthesized Shanghai dialect is rather worse.

The paper is organized as follows: Section~\ref{sec:tts-system} reports our approach to building End-to-end TTS systems for the two tasks.
Section~\ref{sec:evaluation} reports the evaluation results.
Section~\ref{sec:discussion} is our discussion, and we finally conclude in Section~\ref{sec:con}.

\section{TTS system building recipe}~\label{sec:tts-system}
In this section, we start with reporting our data preparation. Since it's 
the first priority for building a successful TTS system, no matter what kind of methods is employed. After that, we introduce our efforts on the three components of a typical TTS system. That is, a front-end component for the linguistic feature generation, 
an End-to-end acoustic models realizing linguistic feature to acoustic feature mapping using Tacotron2,
and a WaveNet Vocoder realizing acoustic feature to waveform generation.

\subsection{Data Processing} \label{sub:data-preces}
In Blizzard Challenge, the data released by the organizer is normally far from clean. Typical issues include: 1) the acoustic utterances are rather long, and some of utterances have poor acoustic conditions; 2) the transcription are not exactly transcribed and yet to be normalized; 3) the linguistic unit for acoustic model training should be obtained with the help of language specific lexicon\footnote{Lexicon is usually necessary for a better TTS system. Even for the state-of-the-art End-to-end TTS method, phone sequence would be better than grapheme sequence yielding better TTS system.}.
\subsubsection{Data preparation}~\label{subsub:data-preparation}
\begin{figure}[th]
    \centering
    \begin{subfigure}[b]{0.23\textwidth}
        \includegraphics[width=\linewidth]{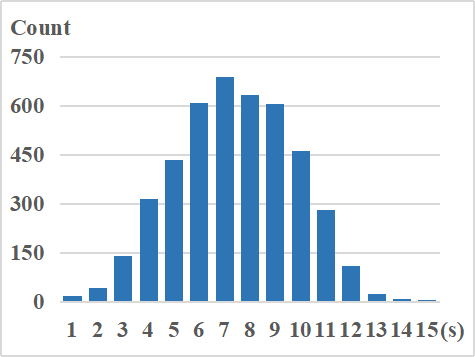}
    \caption[]{Original Mandarin}
    \end{subfigure}
    \begin{subfigure}[b]{0.23\textwidth}
        \includegraphics[width=\linewidth]{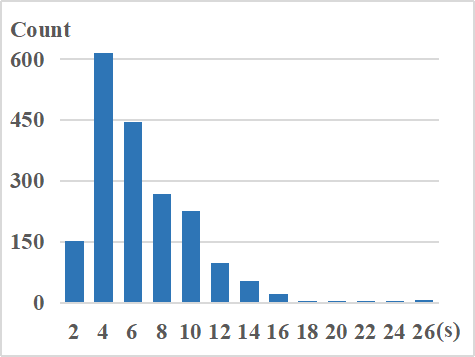}
    \caption[]{Original Shanghai dialect}
    \end{subfigure}
        \begin{subfigure}[b]{0.23\textwidth}
        \includegraphics[width=\linewidth]{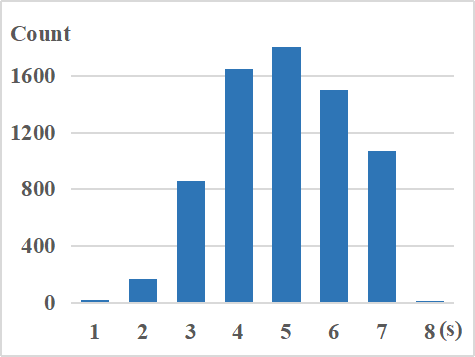}
    \caption[]{Resegmented Mandarin}
    \end{subfigure}
    \begin{subfigure}[b]{0.23\textwidth}
        \includegraphics[width=\linewidth]{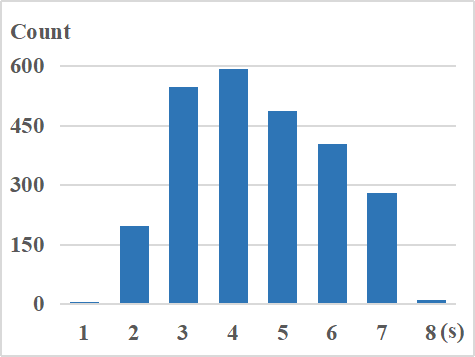}
    \caption[]{Resegmented Shanghai dialect}
    \end{subfigure}
    \caption{ Utterance length distribution before and after utterance re-segmentation for the ease of training.}
    \label{fig:utterance-length-dis}
\end{figure}

For this year data preparation, we first cut the longer utterance into smaller segments according to the silence indication.
This is crucial for End-to-end system training, since we found longer utterance with limited training data can lead to poor acoustic models~(fail to align in the end of utterance, that is, unstoppable repetition or word missing). 
Figure~\ref{fig:utterance-length-dis} reveals our data length distribution before and after utterance segmentation.

As shown in Figure~\ref{fig:utterance-length-dis}, we restrict the maximal utterance length around 7 second for both tasks. After re-segmentation, the length variation of the overall utterance are reduced a lot. Besides,
when segmenting the longer utterance into shorter utterance, we also correct/normalize the transcription simultaneously. After that, we train a Mandarin ASR system with the processed Mandarin data, using Kaldi toolkit~\cite{povey2011kaldi}, and employ such an ASR system to align the training data. We remove those utterances that fails to align. 

\subsubsection{Feature preparation}~\label{subsub:feature}
We have both input and output features to prepare. That is, the front-end linguistic  features that are generated from the normalized text transcription or the phonetic sequence with the help of the lexicon, and the output acoustic features. We choose 80-dimensional Mel scale filter-bank features as our acoustic feature. For the front-end linguistic features, we use the character/letter that is from corresponding Mandarin initial-final-based syllables. Usually, one can embed other linguistic knowledge, such as word features, prosodic features to improve the prosodic expressive capability, as well as intelligibility of the synthesized waveform.

\subsection{End-to-end acoustic Model}~\label{sub:am}
We use Tacotron2~\cite{shen2018natural} as the acoustic model to map syllable-based letter sequence to 80-dimensional Mel filter-bank sequence. The model architecture and the training method are similar to what are proposed in~\cite{shen2018natural}. We use Tacotron2 as our acoustic model because  it is much simpler compared with the WaveNet-based TTS method.

Unlike WaveNet-based TTS method~\cite{oord2016wavenet,shen2018natural,oord2018parallel}, Tacotron cannot generate time-domain-based waveform directly, it's output is a low-level acoustic features, such as short-time Fourier transform (STFT) features, or Mel filter-bank features as employed in our work. The output features can be employed to generate waveform with different means, such as
algorithm-based Griff-Lim method, or conditional  WaveNet-vocoder-based method. Besides, the input of the 
Tacotron method is much simpler than the WaveNet-based TTS method. It only uses pure linguistic features, while the input of the WaveNet-based method not only contains and context-dependent linguistic features, but also contains  acoustic features, such as F0, which can not be reliably obtained.

% Tacotron2 is a typical End-to-end acoustic model structure consisting of an encoder and a decoder.

% The encoder consists of three layers of CNN network and two layers of BLSTM network. Each CNN layer contain 512 filters with shape 5 followed by batch normalization and ReLU activations. The output of the final convolutional layer is passed into the BLSTM layer containing 512 units.
% The decoder is a autoregressive recurrent neural network architecture which contains a small pre-net containing 2 fully connected layers of 256 hidden ReLU units, two layers of the LSTM with 1024 units based on Location Attention, and a Post Net containing five layers of CNN comprised of 512 filters with shape 5x1 followed normalization and tanh activations to predicts a residual.

% We tried different modeling units as the input of the acoustic model, such as phone, syllable, Mandarin Character, but finally selected each character of syllable as the modeling unit, because the particle size of syllable's character is relatively small, we tried to increase The number of CNN layers in the encoder has been increased to

% We use Tacotron2 as the architecture of acoustic model and build by ESPNet toolkit \footnote{ESPNet: https://github.com/espnet/espnet}.
% The input of acoustic model is syllable generated by ASR system.
% The acoustic feature is 80 dimension Fbank extracted by ESPNet.

\subsection{WaveNet Vocoder}~\label{sub:vocoder}
We use conditional WaveNet~\cite{oord2016wavenet} as our vocoder to synthesize the waveform once we obtain the estimated acoustic features from  Tacotron2 model. The conditional features are 80-dimensional Mel filter-bank features as mentioned in Section~\ref{sub:am}, which are actually dependent on the output of the front-end Tacotron model.
Compared with the original work~\cite{oord2016wavenet}, we have the following specific configuration.
We use 24 causal convolutional layers, which means our WaveNet have an intermediate receptive field.
In ~\cite{shen2018natural}, it is shown decent results can be obtained even using 12 causal convolutional layers~(versus original 30 layers of network) that are also conditioned on the 80-dimensional Mel filter-bank features.
Besides, the output distribution of the WaveNet is modelled with 10-mixture of discretized logistic distributions. The output of the WaveNet is 16kHz 16-bit waveform.

To train the WaveNet vocoder, we are using the estimated acoustic features, instead of using the ground-truth features. In~\cite{shen2018natural}, it is shown using ground-truth features yield worse synthesized results due to that the WaveNet trained with the fast variational features cannot generalize.

% We use WaveNet~\cite{oord2016wavenet,shen2018natural,oord2018parallel} as our vocoder and 80-dimensional fbank as WaveNet's condition to synthesize sound. WaveNet is an autoregressive network architecture based on dilated casual convolution. Our Vocoder contains 4 stacks, and each stack contains 6 layers of dilated casual convolution layer, so a total of 24 layers of dilated casual convolution layer are used. The sampling rate of audio is down-sampled to 16K Hz. And the discretized mixture of logistics distribution is used To predict that each quantized number is 65536 samples. In the training phase, we directly use the ground truth acoustic features to train WaveNet. In the synthesis phase, we directly use the acoustic features predicted by tacotron2 as input.

% \subsection{LPCNet Vocoder}
% We also consider LPCNet as a vocoder in the system. LPCNet is a WaveRNN variant that combines linear prediction with recurrent neural networks to significantly improve the efficiency of speech synthesis. There are some others vocoders[8,9,10] that used in TTS system. Actually, we use Tacotron2 architecture to predict 20 features that are consisted of 18 bark-scale[5,6] and 2 pitch parameters from input of token sequences, after that, we synthesis speech from 20 features by LPCNet vocoder.
% Vocoder is a significant part for TTS, In our experiments, we use an efficient neural vocoder called Lpcnet, Lpcnet inference runs faster than realtime on a single CPU while producing a high quality speech output.

\subsection{Model Training}
As mentioned above, our TTS system pipeline contains two main components, namely, a Tacotron2-based acoustic model, and a WaveNet-based vocoder. We use ESPNet toolkit~\cite{hayashi2020espnet} to train  Tacotron2 model, and another open source \texttt{r9yr}~\footnote{https://github.com/r9y9/wavenet\_vocoder} to train WavNet vocoder respectively.

\begin{figure}[th]
\centering
\captionsetup{justification=centering}
\includegraphics[width=\linewidth]{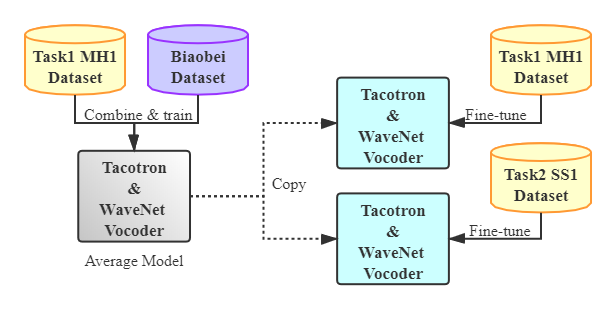}
\caption[]{A simple average speaker modeling approach to low-resource text-to-speech synthesis using Tacotron + WaveNet pipeline}
\label{fig:average-speaker}
\end{figure}

One of the main challenges comes from low-resource issue, since we have only about \texttt{3} hours of Shanghai dialect data for the Shanghai dialect TTS task. Even for the Mandarin task, the organizer only released about 8 hours of training data, it is  not enough to train a robust and higher-quality TTS system. As a result, we think of using more external  data to train an initial average speaker TTS system. After that we are using the target Mandarin and Shanghai dialect data to fine-tune the average speaker models respectively. The whole framework is illustrated in Figure~\ref{fig:average-speaker}.
Since we cannot get Shanghai dialect data,
here we use the \texttt{Biaobei}~\footnote{https://www.data-baker.com/index.html} data set that is public available for Mandarin text-to-speech research work.

From Figure~\ref{fig:average-speaker}, we merge the \texttt{Biaobei} and the 8-hour Mandarin data to train 
both Tacotron2 and WaveNet networks. What we are doing 
is actually a simplified speaker average modeling approach. We use 2 speakers to train the average models, however no speaker dependent embedding is considered.

Besides, we are forcing the Shanghai dialect and the Mandarin acoustic models to share the same phonetic set. As a result,
our acoustic model belong to a monolingual acoustic model. 
Additionally, we are even not including the Shanghai dialect data to train the average models, for fear of our Mandarin  text-to-speech synthesis quality being affected.

However, after results submission, we are aware of such defects as mentioned. Shanghai dialect is greatly different from Mandarin, consequently their phone sets cannot be shared, and the acoustic models should be  multilingual if we want to sufficiently exploit all the data of what we have.

% We used the Mandarin data of Task 1 and the standard female voice data of Biaobei\footnote{Biaobei: https://www.data-baker.com/index.html} open source to build our Mandarin system, and used ESPNet\cite{hayashi2020espnet} to train the tacotron2\cite{shen2018natural} acoustic model. First use the ASR system to get the audio alignment and convert it into Chinese pinyin. Taking Chinese Pinyin as input and 80-dimensional fbank as output, train tacotron2 to get an average model. Then we use the male Chinese Mandarin data provided by task1 to fine-tune the average model to get a more stable acoustic model.

% When training Vocoder, use the 80-dimensional fbank proposed by ESPNet as the condition of WaveNet
% \footnote{With https://github.com/r9y9/wavenet\_vocoder}, and use the competition Mandarin data and standard open source standard female voice to train WaveNet, and use the competition data to fine-tune on this basis to get the final Vocoder of task1.

% In the construction of the task2 Shanghai dialect system, because the Shanghai dialect data is very limited, we try to make the Shanghai dialect and Mandarin use the same phoneme, so that we can do transfer learning on the acoustic model and vocoder of task1 separately. Data to construct the Shanghai dialect system. Using the same phoneme in Shanghai dialect and Mandarin will inevitably cause some mismatches in the synthesis of Shanghai dialect, which will eventually be reflected in the submitted assessment set.

\section{Evaluation results}~\label{sec:evaluation}
\begin{figure}[th]
\centering
\captionsetup{justification=centering}
\includegraphics[width=\linewidth]{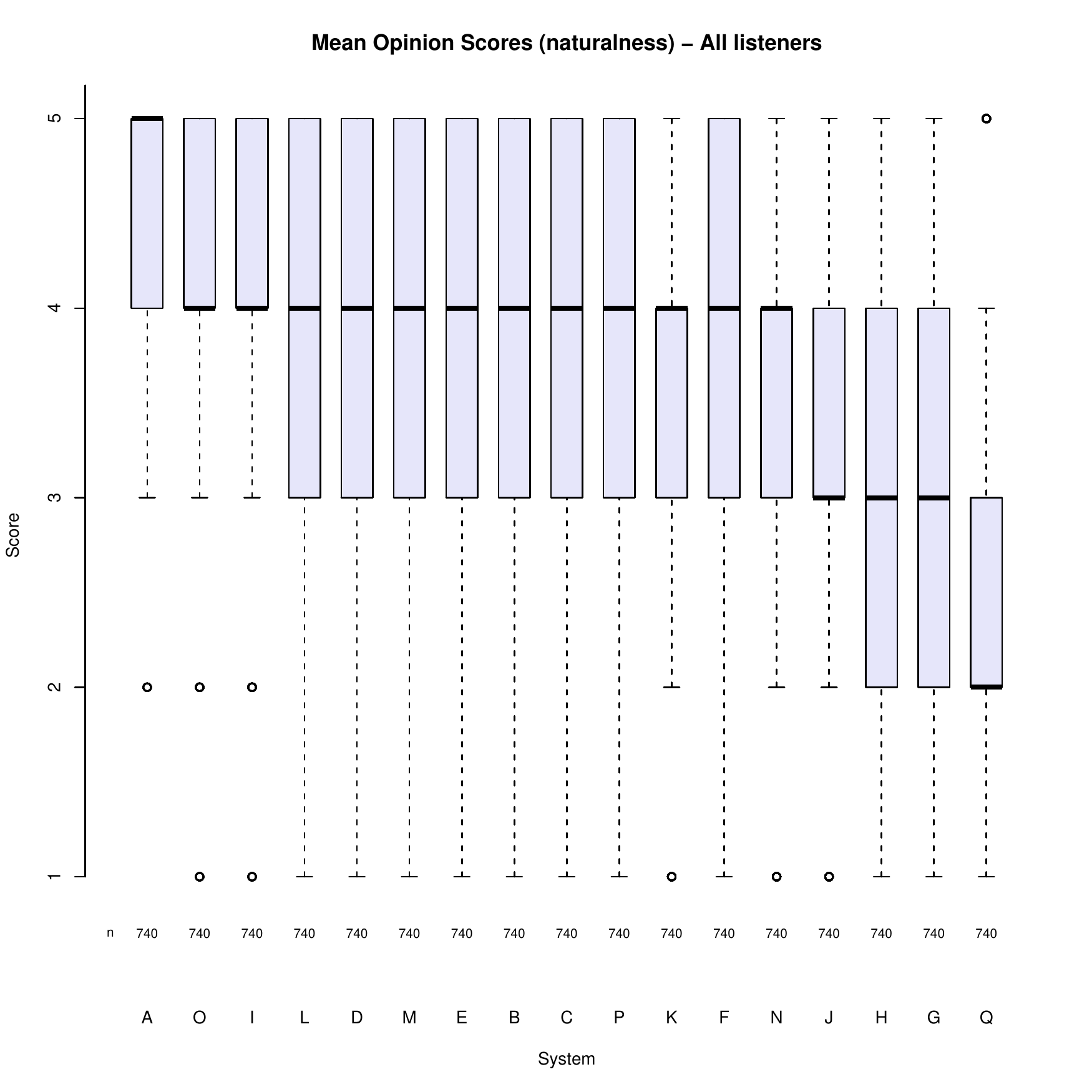}
\caption[]{All MOS for Mandarin Task (MH1)}
\label{fig:mh1-all-mos}
\end{figure}

We report the evaluation results for our Mandarin and Shanghai Dialect text-to-speech systems in this section. the organizer has adopted three main subjective measures to evaluate the overall performance of a submitted system. 
That is, score of Mean Opinion Score (MOS), score of similarity to the original speaker, and score of intelligibility for the synthesized speech. Figure~\ref{fig:mh1-all-mos}
reports the MOS of the Mandarin systems from overall participants, and  our system is  represented as \texttt{K}.

From Figure~\ref{fig:mh1-all-mos}, the MOS score for system \texttt{K} is about 3+ in the range range of [0, 5]. Compared with the other team's performance that are indicated in Figure~\ref{fig:mh1-all-mos}, it
belongs to the third tier though it is not the worst. 
Due to the space limitation reason, we do no report the other two evaluation scores for our Mandarin TTS system, since they in a similar rank as is revealed in Figure~\ref{fig:mh1-all-mos}.

\begin{figure}[th]
\centering
\captionsetup{justification=centering}
\includegraphics[width=0.8\linewidth]{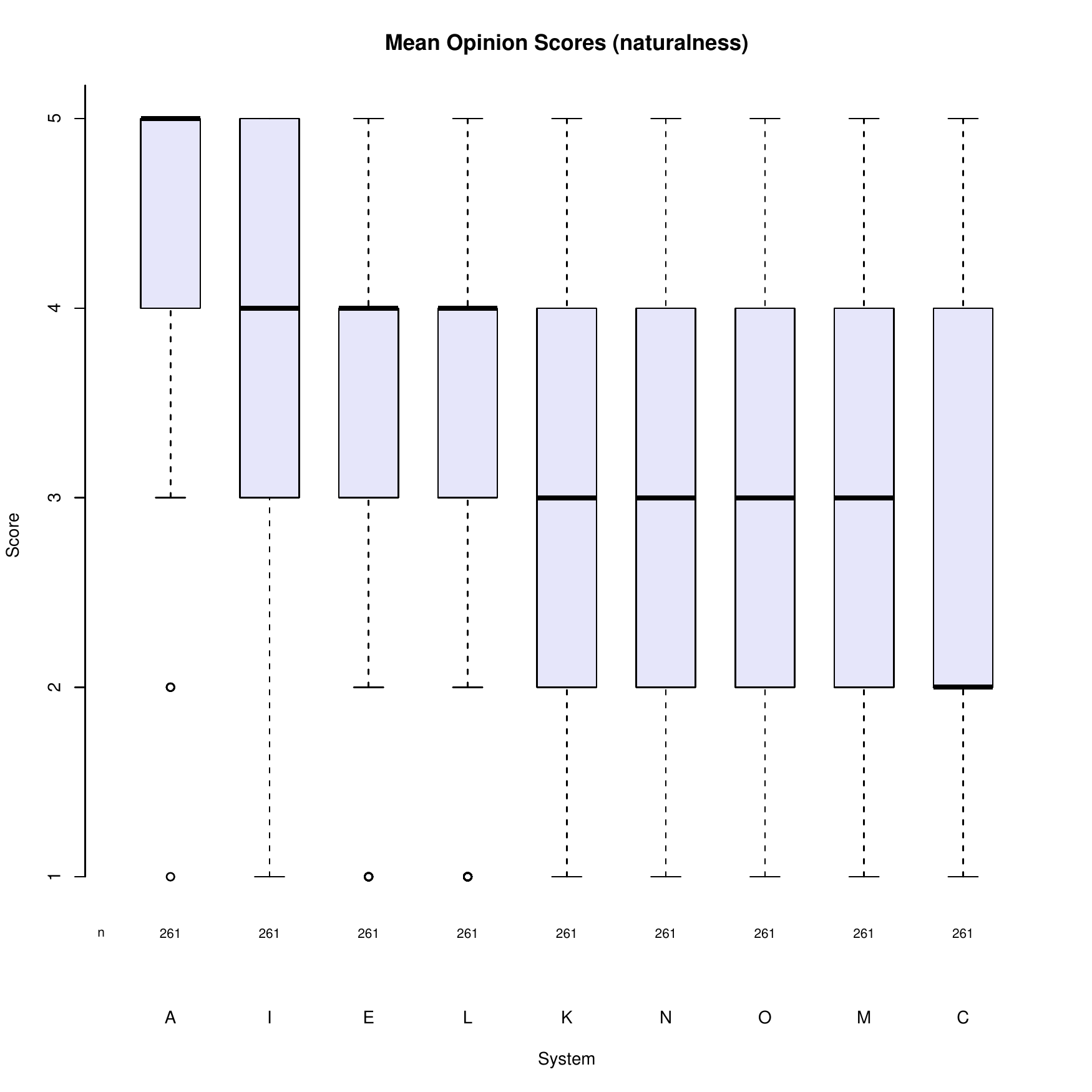}
\caption[]{All MOS for Shanghai Dialect task (SS1)}
\label{fig:ss1-all-mos}
\end{figure}

Figure~\ref{fig:ss1-all-mos} reports the  MOS performance of the overall Shanghai dialect systems, of which our entry system is also represented as \texttt{K}.
From Figure~\ref{fig:ss1-all-mos}, our Shanghai dialect MOS score is about 3, which are worse than the one that is obtained for our Mandarin system in Figure~\ref{fig:mh1-all-mos}. Overall speaking, the MOS score of system \texttt{K} are ranked at middle level. However,
compared with Figure~\ref{fig:mh1-all-mos},  what is reflected in Figure~\ref{fig:mh1-all-mos} suggests there are fewer participants~(or valid entry systems) for the Shanghai Dialect text-to-speech task, and the  MOS performance for the overall entry systems  are worse. 
Such a performance degradation might attribute to less training data being available.

\begin{figure}[th]
\centering
\captionsetup{justification=centering}
\includegraphics[width=0.8\linewidth]{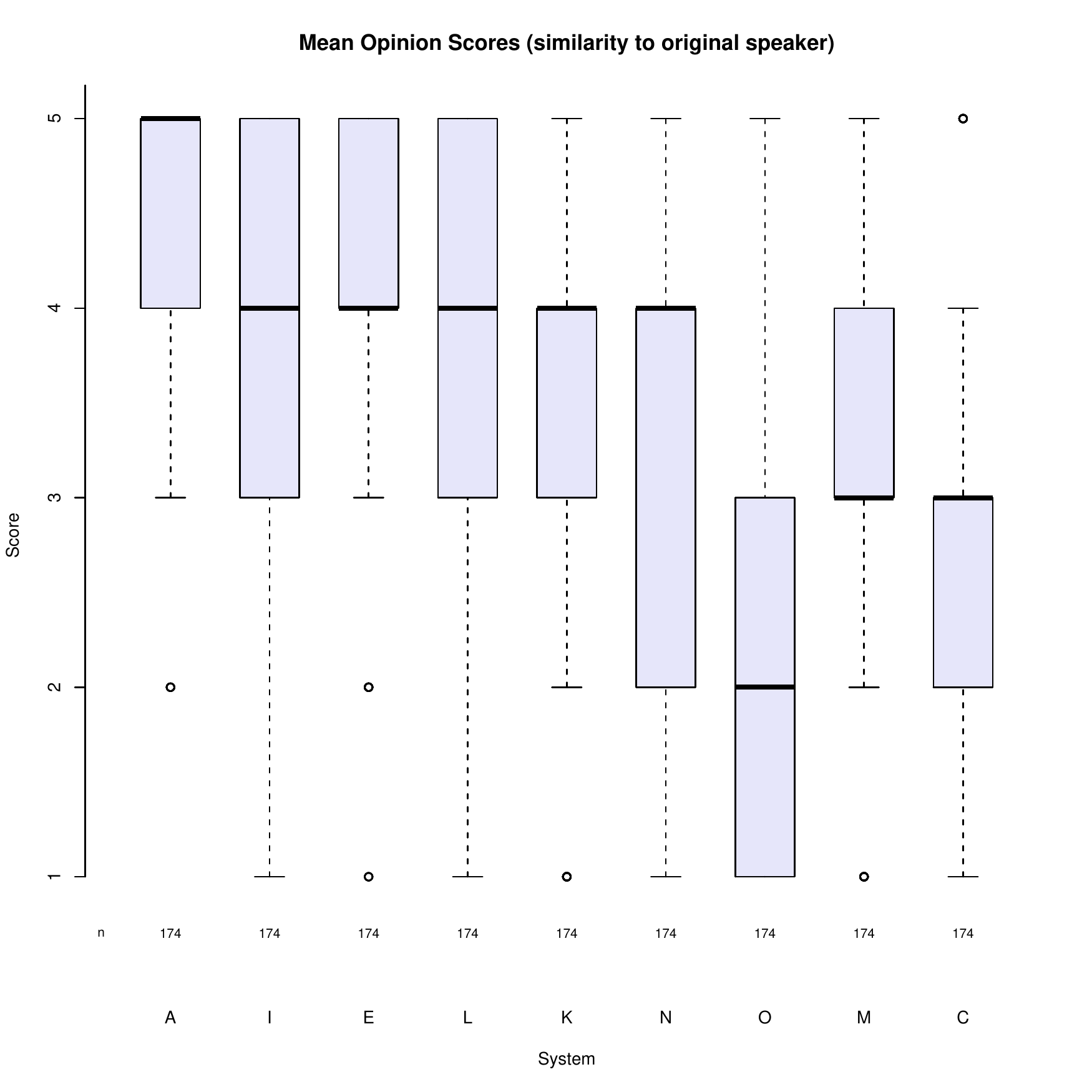}
\caption[]{Similarity to original speaker for Shanghai Dialect task (SS1)}
\label{fig:ss1-all-sim}
\end{figure}

Figure~\ref{fig:ss1-all-sim} shows the scores of the similarity  to the original speaker for  all the participants
in Shanghai dialect text-to-speech task. the score of the system \texttt{K}  is around 3+, and is again positioned at middle level in the overall participants.

\begin{figure}[th]
\centering
\captionsetup{justification=centering}
\includegraphics[width=0.8\linewidth]{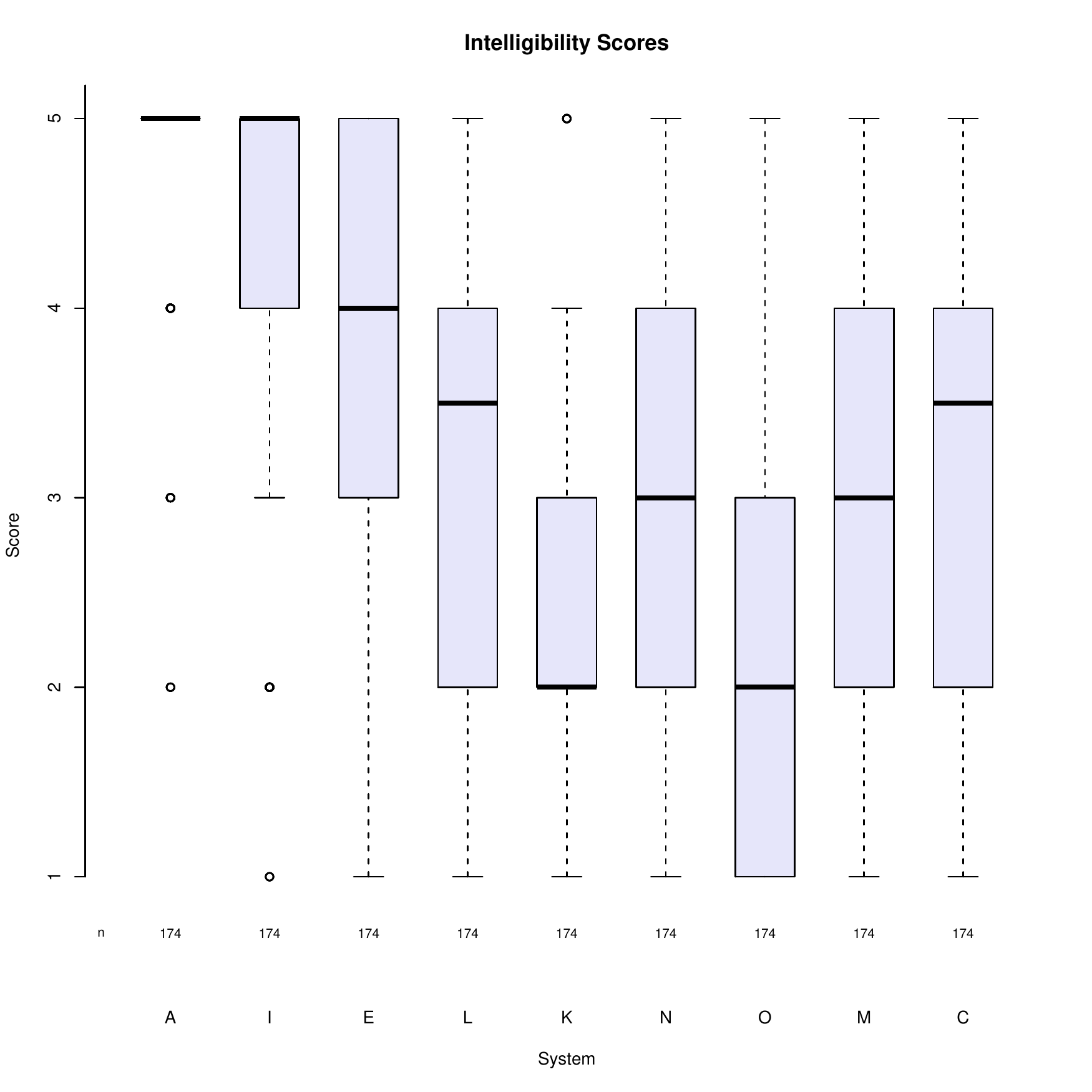}
\caption[]{All Intelligibility Scores for Shanghai Dialect task (SS1)}
\label{fig:ss1-all-int}
\end{figure}

Figure~\ref{fig:ss1-all-int} presents intelligibility scores of all the participants.
Our intelligibility score is almost the worst. This is not surprising. Since we assume
we have zero knowledge about Shanghai dialect in terms of word pronunciation, we 
take Mandarin word pronunciation instead. This pronunciation sharing contradicts the 
facts that Shanghai dialect has completely different pronunciation from Mandarin pronunciation.
After system submission, we are aware we have two choices for better results. One is 
to let Mandarin and Shanghai dialect share the same formality of pronunciation rules, but 
with different phone sets. That is, we append the letter from the same syllable with different language
identifiers. This should make a difference but not make too much sense.
The other choice is to directly use Shanghai-dialect-based syllable as provided by the
organizer. Either choice belongs to a multilingual acoustic modeling approach if we merge the two data sets to train the system. 

\section{Discussion}~\label{sec:discussion}
As is shown in Section~\ref{sec:evaluation}, our performance are left behind by the top tiers. In this section, we are meant to  analyze what remains to be done in ahead of time.

First, we think it is important to train a multilingual acoustic model in future.
This is especially important for low resource language. The effectiveness of multilingual training has already proved in automatic speech recognition area~\cite{kannan2019large}.
As is revealed in Figure~\ref{fig:ss1-all-int}, the intelligibility of the Shanghai Dialect is much worse than that of other competing systems.
This is because we  made a mistake by using Mandarin phone set for Shanghai Dialect, which is obviously different from the former in terms of word pronunciation though the written form of the two languages are the same. 
Actually, if we use letter-based  context-independent unit, 
multilingual acoustic model training is completely feasible. This is because
a given language usually contains only dozens of language-dependent letters/characters that are from its phone set which itself is also far below 100.

Secondly, our front-end linguistic features are too simple; we only have letter embedding features as input.
This might not be a severe issue if one has enough training data. However, for a low-resource text-to-speech system development. High-level features, such as word and prosodic features~\cite{tian2019tencent}, could benefit a lot.
For instance, we can employ off-the-shelf BERT system to generate word embedding features to boost our encoder for the Tacotron models.
Similarly, we can also add prosodic features by means of using Conditional Random Fields~(CRF)~\cite{xiao2020bertch} to label the input text utterance. All these are worth our efforts in future.

Thirdly, although using WaveNet as neural vocoder can yield very desirable synthesized speech, it's computational cost is huge, and not applicable in real case. During system training, we also attempted to use other neural vocoders. Particularly, we tried LPCNet~\cite{valin2019lpcnet}. We found LPCNet can also synthesize good voice, however it seems to yield lower-quality speech compared with WaveNet.

\section{Conclusions}~\label{sec:con}
We reported our work for the NTU-AISG text-to-speech entry system in Blizzard Challenge 2020. We have participated two tasks. One is a Mandarin TTS task and the other is a Shanghai Dialect TTS task. To address the low-resource issue, we employed a simplified 
 average speaker modeling method for both Tacotron2-based acoustic model and 
WaveNet-based vocoder. However, we forced Mandarin and Shanghai dialect to share the same syllable set, such an assumption contradicts the real linguistic facts.
 As a result, our text-to-speech system for the Shanghai dialect task yields poor intelligibility. 
 In future, we are conducting real average speaker modeling method, as well as multilingual training research.
% \section{Acknowledgements}

\bibliographystyle{IEEEtran}
\clearpage
\bibliography{mybib}

\end{document}